\documentstyle[11pt]{article}

\font\tenrm=cmr10

\newcommand{\be}{\begin{equation}}
\newcommand{\ee}{\end{equation}}   
\newcommand{\swch}{Schwarzschild}
\newcommand{\veff}{V^{\rm eff}}

\title
{\huge  Canonical Quantization Inside the
   Schwarzschild Black Hole}
\author{U. A. Yajnik\thanks{yajnik@niharika.phy.iitb.ernet.in}\quad
  and\quad  K. Narayan\thanks{narayan@tristan.tn.cornell.edu 
(after September 1 1997)}\\
{\tenrm Physics Department, Indian Institute of Technology Bombay,}\\
{\tenrm Mumbai 400\thinspace076}}
\date{{\small [gr-qc/9706021]}} 
\begin{document}
\maketitle
\begin{abstract}
{We propose a scheme for quantizing a scalar field over the
Schwarzschild manifold including the interior of the horizon.
On the exterior, the timelike Killing vector and on the horizon
the isometry corresponding to restricted Lorentz boosts can be
used to enforce the spectral condition. For the interior we
appeal to CPT invariance to construct an explicitly
positive definite operator which allows identification of
positive and negative frequencies. This operator is the
translation operator corresponding to the inexorable propagation
to smaller radii as expected from the classical metric. We also
propose an expression for the propagator in the interior and
express it as a mode sum. The field theory thus obtained is
meaningful for small curvatures far from the classical singularity.}
\end{abstract}
\vskip 2cm
\begin {center}
{\sl To appear in Classical and Quantum Gravity}
\end{center}

\newpage
\section{Introduction}

\par
We propose a quantization of a scalar field inside the Schwarzschild
black hole. We adopt the  conventional canonical
quantization approach and the well known
classical fact that in the \swch\ interior test particles move
inexorably inward. It thus seems reasonable to assume that in
the quantum theory we take the radial momentum operator to be
{\it positive} definite in the direction of {\it decreasing} $r$
co-ordinate.  However, the inclusion of out-going modes with
positive outward radial wave-number $k_{\omega}$ would also be
required by completeness.  These then have to be interpreted as
in-going modes with positive inward wave-number but of opposite
charge. This generalises the CPT theorem as usually formulated
in Minkowski space. There is a well known lucid explanation for
the existence of antiparticles in Minkowski space due to Feynman;
an interesting thought experiment was given by Weinberg \cite{wein}.
Here we  obtain a generalization to the black hole interior.

From the axiomatic point of view the main issue is the spectrum
condition, viz., the existence of a positive definite operator
whose spectrum would guarantee the existence of a ground state.
For the case of curved spacetime, Haag et al \cite{haag1} have
formulated this as the principle of local 
stability of the Wightman functions. It requires that
the support of the two-point function be restricted to the
forward, i.e. here, the radially inward lightcone.
We show that this principle holds for our quantization.

Candelas and Jensen \cite{jencan} extended the Feynman Green
Function with Hartle-Hawking boundary conditions to the interior
of the Schwarzschild black hole. This Green Function obeys a
periodic boundary condition in the time co-ordinate, and thus
assumes a Kubo-Martin-Schwinger type state as the state of
lowest energy. This is 
appropriate to the presence of an asymptotic heat bath.
The boundary conditions that our prescription entails are 
similar to those of Boulware\cite{Boul}
who has determined the vacuum state for the global Schwarzschild
manifold. We provide an {\it a priori} motivation for the vacuum 
state strictly in the interior and construct a causal propagator. 
Our results agree with those of Boulware in the interior. 
However, our prescription leaves open the possibility of
matching with different quantizations in the exterior.

This paper is organised as follows. In section 2 we begin with
quantization conditions and the re-interpretation of the modes,
in section 3 we introduce the radial momentum operator. In
section 4 we construct the causal propagator.  Section 5 contains
discussion and outlook.

\section{Quantization Inside the Schwarzschild Black Hole}
       
\par  
   The Schwarzschild spacetime is described by the spherically
   symmetric line element

\be
\label{lineelem}
   ds^{2} = (1 - \frac{2m}{r}) dt^{2} - \frac{dr^{2}}{1 - \frac{2m}{r}} 
           - r^{2} d{\theta}^{2} - r^{2}\sin^{2}\theta {d\phi}^{2}       
\ee
in the $\{t,r,\theta,\phi\}$ coordinate system. 
These coordinates permit separation of variables and are
suitable for finding the mode functions.
We shall also need to use the so called tortoise co-ordinate\cite{mtw}
$r^*$$=$$r+2m\ln|r/2m-1|$ in which the metric is
\be
\label{linestar}
ds^{2} = (1 - \frac{2m}{r}) (dt^{2} - dr^{*2}) 
           - r^{2} d{\theta}^{2} - r^{2}\sin^{2}\theta {d\phi}^{2}       
\ee
In this form the $t-r$ part is conformal to 1+1 Minkowski space. 
For $r<2m$, $r^*$ ranges from $0$ to $-\infty$ as a
monotonically decreasing function of $r$. The
spacetime has the Killing symmetries corresponding to time translation 
invariance $(t \rightarrow t + \tau)$ and rotational invariance. It
has an event horizon at $r=r_{H}=2m$. 
The Killing vector $\partial/\partial t$ is timelike in the exterior, null
on the horizon and spacelike in the interior.  The vector
$\partial/\partial r$ is spacelike in the exterior and timelike in the
interior.  Thus, inside the black hole, the roles of time and space are
reversed.  Furthermore, any particle inside the black hole is
inexorably drawn to the singularity. In other words, the future
lightcone of a particle in the interior does not cross the horizon and
necessarily terminates at the singularity.  This is what motivates our
prescription for quantization. For simplicity we consider a charged
massless minimally coupled scalar field. The scheme however can be
extended to any realistic field.

Accordingly, defining the canonically conjugate momentum
\be
\label{canmom}
\Pi_r = \frac{\partial L}{\partial(\partial_{r^*} \phi)} = \partial_{r^*} \phi^{\dag}, 
\ee
we impose the radial quantization conditions
\be
\label{qcondr1}
   [\phi(r,t,\Omega),\frac{\partial
\phi^{\dag}}{\partial r^{*}}(r,t',\Omega')] = 
i \delta_{\Sigma}(t,\Omega;t',\Omega')
= i [-g(\Sigma)]^{-1/2} \ \delta(t - t') \ \delta(\Omega - \Omega')
\ee
\be
\label{qcondr2}
[\phi(r,t,\Omega),
   \phi^{\dag}(r,t',\Omega')] = 0 =
[\frac{\partial}{\partial r^{*}}\phi(r,t,\Omega),
\frac{\partial}{\partial r^{*}} 
   \phi^{\dag}(r,t',\Omega')] 
\ee 
These are ``equal $r$ commutation relations" for the field.
Quantization is performed on constant-$r$ hypersurfaces near the
horizon in the interior (denoted by $\Sigma$), which are Cauchy
surfaces in the black hole interior.
Consider the local stability requirement of \cite{haag1}. Paraphrased for the
present case, it requires 
   that the 2-point function $W^{(2)}(z_{1},z_{2})$ has support on the
   forward, i.e., radially $inward$ light cone 
\be   
p \cdot p \geq 0
\ee
with the radial   component of the momentum 
\be
p^r \leq  0   
\ee
Our quantization conditions (\ref{qcondr2}) are that $\phi$ and
separately $\pi$ at same $r$ commute whereas according to
(\ref{qcondr1}) $\phi$-$\pi$ commutator is nontrivial in the inward
directed light-cone. Thus these can be taken to be implementation of
the  local stability requirement. 

Minkowski space quantization on hypersurfaces of constant $t$ is covariant
under rigid Lorentz transformations of the cartesian $t-\vec x$ 
coordinates. The background geometry 
of the blackhole in \swch\  co-ordinates suggests a preferred slicing for
quantiztion as chosen above. The hypersurfaces of constant $r$ 
are also those with ${\rm Tr}K={\rm constant}$ where $K$ is the extrinsic 
curvature of the hypersurface. This choice of foliation is standard in
dynamical formulation of gravity \cite{wijaw}.  
Here the resulting field theory 
will be covariant under constant translations of the $t$ co-ordinate and
rigid rotations centred on the origin already chosen. 
A quantization scheme based on any other co-ordinates will 
be inequivalent to the present one. 

We now proceed to obtain a representation of this algebra by
introducing creation and annihilation operators.
Consider the  massless scalar wave equation  $\Box \phi = 0$
with $\Box$ the appropriate d'Alembertian operator for the black hole
interior. Separating this in
the $\{t,r,\theta,\phi\}$ coordinates, we obtain the mode functions

\begin{equation}\label{modefunc}
h_{\omega lm}(r,t,\Omega) \sim  \frac{R_{\omega l}(r)}{r} \ Y_{lm}(\theta
,\phi) \ e^{-i\omega t} 
\end{equation}
where $Y_{lm}$ are the spherical harmonics. In the co-ordinate $r^*$,
 the equation satisfied by $R_{\omega l}$ is 

\begin{equation}
\label{reqn}
\frac{d^{2}R_{\omega l}}{dr^{*2}} + (\omega^{2} - [\frac{l(l+1)}{r^{2}}
+ \frac{2m}{r^{3}}][1 - \frac{2m}{r}])R_{\omega l} = 0
\end{equation}
Writing this in the form       

\begin{equation}
\frac{d^{2}R_{\omega l}}{dr^{*2}} + k_{\omega}^{2} R_{\omega l} = 0
\end{equation}       
we can identify the positive definite quantity 

\begin{eqnarray}
k_{\omega}^{2} &=&  \omega^{2} - V^{\rm eff} \label{veffdef}\\
          &\equiv&   \omega^{2} - [\frac{l(l+1)}{r^{2}} 
		+ \frac{2m}{r^{3}}][1 - \frac{2m}{r}] \label{krdef}
\end{eqnarray}
with the radial wave-number squared of the mode. The $\veff$ vanishes
at the horizon so that $k_{\omega}=\omega$ there. Further, close to the 
horizon, i.e., for $2m-r\ll 2m$, $R_{\omega l} \sim e^{\pm i(k_{\omega}r^{*} -
\omega t)}$. 
We shall now on take $k_{\omega}$ to be the positive square root of the
above
equation, and choose $h_{\omega lm}$ to be those mode functions which
satisfy near the horizon\footnote{The $e^{-i\omega t}\psi^l(r,-\omega)$ 
of \cite{Boul} correspond to our $h^*_{\omega lm}$} 
\begin{eqnarray}
\frac{\partial}{\partial r^{*}} h_{\omega lm}(r,t,\Omega) &=& -i
   k_{\omega} \ h_{\omega lm}(r,t,\Omega) \nonumber \\
&\phantom{=}&\hskip 20 pt ({\rm for}\quad r<2m\quad {\rm and}\quad 2m-r\ll
2m) \label{posmode}
\end{eqnarray}
i.e., the positive definite eigenvalues are associated with {\it
ingoing} radial momentum. This is analogous to the condition for 
positive frequency modes in Minkowski space.
With appropriate normalisation, these modes satisfy the completeness
relation 
\be
\label{comprln}
   \sum_{lm} \int d\omega \ h_{\omega lm}(r,t,\Omega) \ h_{\omega
   lm}(r,t',\Omega') = \delta_{\Sigma}(r,t,\Omega;r,t',\Omega')
= \frac{\delta(t - t') \ \delta(\Omega - \Omega')}{r^2}
\ee
where $\Sigma$ is the surface of quantization, a constant-$r$
hypersurface near the horizon, as before.
In the mode expansion we now take summation over modes with
the parameter $\omega$ taking both positive and negative values while
$k_{\omega}$ remains positive. Accordingly, the interior Fourier
expansion near the horizon is
\be
\label{modeexp}
\phi_{in} = \sum_{lm} \int^{\infty}_{-\infty} 
\frac{d\omega}{\sqrt{2\pi} \thinspace r\thinspace \sqrt{2k_{\omega}} }
   (a_{\omega} \ e^{-i(k_{\omega}r^{*} - \omega t)} Y_{lm} +
	b_{\omega}^{\dag} \ e^{i(k_{\omega}r^{*} - \omega t)} Y^{*}_{lm})
\ee
where we have written $h_{\omega} \sim e^{-i(k_{\omega}r^{*} - \omega
t)}/r$ as
   asymptotic forms near the horizon of general modes $h_{\omega lm}$.
The indices $l,m$ on $a_{\omega}$ and $b_{\omega}$ have been suppressed.
Now imposing the quantization conditions (\ref{qcondr1}),
(\ref{qcondr2}) we obtain for the expansion parameters,

\be       
\label{qcondab}
[a_{\omega}, a_{\omega'}^{\dag}] = \delta(\omega - \omega')
   = [b_{\omega}, b_{\omega'}^{\dag}] 
\ee
and that all other commutators vanish.

We can justify the interpretation of modes implied by conditions
(\ref{qcondab}) by paraphrasing the CPT invariance argument of Weinberg 
\cite{wein}. Consider an observer in the interior who performs an
experiment in which a particle is created at point A and is ``later",
i.e., further down the radial direction is destroyed at B. This
radial (temporal in the interior)
ordering of events cannot be reversed for any classical events by
another observer without a superluminal Lorentz transformation.  However
for those events separated by spacetime intervals less than the Compton
wavelength of the particle, this is not guaranteed. In this case it
is possible to observe a particle of the same charge being 
annihilated at B before being created at A, thus travelling backward in $r$.
The only way causality can
be maintained is to insist that a particle of opposite charge travelled
forward in $r$, emitted from B and absorbed at A. Eq.s (\ref{qcondab}) are
consistent with this interpretation of the modes.

\section{Recovering the Spectrum condition and CPT}
Next we need to verify that the algebra of the operators corresponding to
spacetime symmetries is realised, and in particular that a positive 
definite operator exists, whose spectrum guarantees the existence of 
a ground state upon which the spectrum generated by the creation operators
can be built. To begin with we demand the existence of a vacuum or   
no-particle state $\mid 0^I \rangle$ satisfying 
\be
\label{gstate}
a_{\omega} \mid 0^{I} \rangle = b_{\omega} \mid 0^{I} \rangle = 0,
\ee
The global Killing symmetries of
   time translation (spacelike in the interior) and rotations
   on surfaces of constant $r$ are implemented by the operators

\be
K_{t} = \int d \Lambda \ T^{t}_{t}
\ee
\be
K_{\theta} = \int d \Lambda \ T^{\theta}_{\theta}
\ee
\be
K_{\varphi} = \int d \Lambda \ T^{\varphi}_{\varphi}
\ee       
with $d\Lambda$ denoting the induced 3-volume on constant $r$
hypersurfaces and
with $T$ components obtained in the usual way from the matter Lagrangian.
    These can be thought of as the infalling mass-energy at any
   fixed radius and angular momentum content respectively of the
   quantum field $\phi$. The energy now is not positive definite,
   but this is the operator corresponding to the energy operator
   outside. In the vacuum introduced above, we find 
\be
\langle 0^{I} \mid K_{t} \mid 0^{I} \rangle = \langle 0^{I} \mid K_{\theta} \mid 0^{I} \rangle
   = \langle 0^{I} \mid K_{\varphi} \mid 0^{I} \rangle = 0
\ee
   The above vanishing of the expectation values is clear
because $\int^{\infty}_{-\infty} d\omega \ \omega \ldots$ and
$\sum^{l}_{m=-l}$ etc. appear.
The quantum dynamics is now generated by the operator 
  
\be
K_{r} =  \int d \Lambda \ T^{r}_{r}
\ee
The radial momentum density is

\be
T^{r}_{r} = \Pi_{r} \partial_{r^*} \phi + \Pi^{\dag}_{r}
\partial_{r^*} \phi^{\dag} - L
\ee
\be
\ \ = \frac{1}{\frac{2m}{r} - 1} [\partial_{r^*} \phi^{\dag}
\ \partial_{r^*} \phi + \dot{\phi}^{\dag} \ \dot{\phi} ]
\ee
This is clearly positive definite in the black hole interior.       
Promoted to a quantum operator, $K_r$ is thus positive definite.
Substituting the mode expansion (\ref
{modeexp}) and using the expression for the volume element on
the constant $r$ hypersurface near the horizon

\be
d\Lambda = (2m/r - 1)^{1/2} r^2 \ \sin\theta \ dt \ d\theta \
d\varphi 
\ee
we obtain the following expression for normal ordered $K_r$

\be 
:K_r:\ =\  r (\frac{2m}{r} - 1)^{-1/2} \int^{\infty}_{-\infty}
d \omega \ k_{\omega} \ (a^{\dag}_{\omega} a_{\omega} + b^{\dag}_{\omega}
b_{\omega})
\ee
This makes the state defined through eq.s (\ref{gstate}) a genuine 
ground state as well as a no-particle state. Note that the sign in
front of $k_{\omega}$ in above eqn. is due to our choice the sign of
$k_{\omega}>0$ below eqn. (\ref{krdef}). 

The ground state thus characterised has been shown to be 
stable\cite{Boul}. One may think of it as the 
adiabatic vacuum useful to a freely infalling observer.
The latter should detect particle production, and the same will
be finite if the vacuum specified here is used as the template
with which to compare his local ground states. 
This is similar to what happens in
Friedmann-Robertson-Walker geometries which have a conformal timelike
Killing vector. In the present case we expect copious particle
production as the singularity is approached, much as in the collapsing
phase of relevant FRW metrics. The normal ordering prescription
used may seem arbitrary. But the effect of the infinite
contribution discarded manifests itself as higher derivative
terms in the effective action for gravity. It is possible to choose a
renormalisation prescription such that
when one returns to the gauge specified, the numerical values of
the renormalised operators will be the same as that obtained by
simple normal ordering.

Finally, CPT invariance can be realised by requiring 
\be
C \phi C^{\dag} = \phi^{\dag}       
\ee
which implies the effect of $C$ is
\be
a_{\omega} \leftrightarrow  b_{\omega} 
\ee

Similarly, a restricted parity operator corresponding to reflection 
in the blackhole origin is given by
\be
P^{\Omega} \phi(r,t,\theta,\varphi) {P^{\Omega}}^{\dag} = 
\phi(r,t,\pi-\theta,\varphi+\pi)       
\ee
which implies, under $P^{\Omega}$
\be
a_{\omega,l,m}  \leftrightarrow  a_{\omega l,-m} (-1)^{l}
\ee
and for vector operators
\be
P^{\Omega} V^{\mu}(r,t,\theta,\varphi) {P^{\Omega}}^{\dag} = 
V_{\mu}(r,t,\pi-\theta,\varphi+\pi)       
\ee
The $P^\Omega$  symmetry is complemented by reversal symmetry  of the
$t$ co-ordinate 
\be
\tilde{T} \phi(r,t,\Omega) {\tilde{T}}^{\dag} = \phi(r,-t,\Omega) 
\ee
such that $\tilde{T}$ is unitary. This gives 
\be
a_{\omega} \leftrightarrow a_{-\omega} {\rm \ and \ \ } 
b_{\omega} \leftrightarrow  b_{-\omega}
\ee   
There is no global symmetry reversing the arrow of dynamical 
evolution since $r$-translations are not a symmetry. However 
in the adiabatic approximation, a local and antiunitary $r$ reversal 
operator 
$\tilde{R}(r)$ may be assumed to exist such that for small
increments $\Delta r$,
\be
\tilde{R}(r) \phi(r+\Delta r,t,\Omega) {\tilde{R}(r)}^{\dag} =
\phi(r-\Delta r,t,\Omega) 
\ee

\section{The propagator}

       Since the inexorable direction of propagation is along the
inward radial vector (corresponding to the $t$ variable in the
exterior), the propagator in the interior is expected to be
$r$-ordered. This is analogous to $t$-ordering in the exterior.

\noindent Such a propagator can be written as

\be
iG(x,x') = \ <0^{I}| R \phi(x) \phi^{\dag}(x') |0^{I}> 
\ee
\be
iG(x,x') = \theta(r^* - r'^{*}) <0^{I}|\phi(x) \phi^{\dag}(x') |0^{I}>
\ + \ \theta(r'^{*} - r^*) <0^{I}| \phi^{\dag}(x') \phi(x) |0^{I}>
\ee
The $\theta$ function defined above is the usual step function.
This propagator satisfies the equation

\be
\Box_{x} G(x,x') = \delta(x,x') =[-g(x)]^{-1/2} \delta^{4} (x - x')
\ee
thus making $G(x,x')$ a Green's function for the wave equation
in the Schwarzschild interior geometry.

Near the horizon, the field can be expanded as in eqn.(\ref{modeexp}). 
Using the integral representation for the theta function, the expression 
for $G(r^*,t;{r^*}',t')$ (suppressing the angular co-ordinates) becomes

\be
G(r^*,t;{r^*}',t') =  \lim_{\epsilon \rightarrow 0+} \int^{\infty}_{-\infty}
\frac{d\omega \ d\Lambda}{(2 \pi)^{2} \ r \ r'} \frac{ e^{-i \Lambda (r^{*} -
r'^{*}) + i \omega (t - t')}}{\Lambda^{2} - k_{\omega}^{2} + i
\epsilon} 
\ee

\section{Conclusion}

We have shown that taking account of the classical inexorably
inward motion and incorporating the requirements of CPT, a
unique quantization can be obtained for a matter field in the
black hole interior. The quantization is of the kind possible in
spaces with a conformal Killing vector such as FRW universes. It
can be taken to be the QFT set up by the freely infalling
observer, the equivalent of a comoving observer in FRW
spacetime. We have also shown that imposing microcausality
dictates the form of the propagator at least in the co-ordinates
used. The quantization is expected to break down close to
the classical singularity where semiclassical techniques fail. 
It would be interesting to ask whether there is any signature of
the Hawking radiation detected in this quantization. This
requires matching of this QFT to the standard QFT in the exterior.


\vspace{2cm}

\begin{thebibliography}{9}
\bibitem{wein} S. Weinberg {\sl Gravitation and Cosmology} John Wiley (1973),
sec. 2.13.
\bibitem{haag1} R. Haag, H. Narnhofer, U. Stein, {\sl Comm. Math. Phy.} {\bf
94}, 219 (1984). 
\bibitem{jencan} P. Candelas, B. P. Jensen, {\sl Phys. Rev.} {\bf D33},
1596(1986); see also {\it ibid} pg. 1590.
\bibitem{Boul} D. G. Boulware {\sl Phys. Rev.} {\bf D11},
1404 (1975). Our labelling of mode functions differs 
from this reference. We treat $r^*$ as the natural co-ordinate and
make the split according to positive and negative $k_\omega$.
See our footnote regarding eqn.(\ref{posmode}).
\bibitem{mtw} C. W. Misner, K. S. Thorne and J. A. Wheeler {\sl
Gravitation} W. H. Freeman (1973), pg. 663 
\bibitem{wijaw} see for example J. Isenberg and J. A. Wheeler in {\sl
Relativity, Quanta and Cosmology in the Development of the
Scientific Thought of Albert Einstein}, M. Pantaleo and F.
deFinis, ed.s, vol. I, pp 267-293 (1979)
\end{thebibliography}
\end{document}